# Asymmetrical contact scaling and measurements in MoS$_2$ FETs

Zhihui Cheng[1,2*], Jonathan Backman[3], Huairuo Zhang[4], Hattan Abuzaid[1], Guoqing Li[5], Yifei Yu[5], Linyou Cao[5], Albert V. Davydov[4], Mathieu Luisier[3], Curt A. Richter[2*], Aaron D. Franklin[1,6*]

[1]*Duke University, Department of Electrical & Computer Engineering, Durham, NC 27708, USA*
[2]*National Institute of Standards and Technology, Nanoscale Device Characterization Division, Gaithersburg, MD 20899, USA*
[3]*ETH Zurich, Integrated Systems Laboratory, Zurich, Switzerland, CH-8092*
[4]*National Institute of Standards and Technology, Materials Science and Engineering Division, Gaithersburg, MD 20899, USA*
[5]*Department of Materials Science and Engineering, North Carolina State University, Raleigh, NC, 27695, USA*
[6]*Duke University, Department of Chemistry, Durham, NC 27708, USA*
Email: zhihui.cheng@alumni.duke.edu, curt.richter@nist.gov, aaron.franklin@duke.edu

ABSTRACT

Two-dimensional (2D) materials have great potential for use in future electronics due to their atomically thin nature, which withstands short channel effects and thus enables better scalability. Device scaling is the process of reducing all device dimensions to achieve higher device density in a certain chip area. For 2D materials-based transistors, both the channel and contact scalability must be investigated. The channel scalability of 2D materials has been thoroughly investigated, confirming their resilience to short-channel effects. However, systematic studies on contact scalability remain rare, and the current understanding of contact scaling in 2D FET is inconsistent and oversimplified. Here we combine physically scaled contacts and asymmetrical contact measurements to investigate the contact scaling behavior in 2D field-effect transistors (FETs). The asymmetrical contact measurements directly compare electron injection with different contact lengths while using the exact same channel, eliminating channel-to-channel variations. Compared to devices with long contact lengths, devices with short contact lengths (scaled contacts) exhibit larger variations, 15% lower drain currents at high drain-source voltages, and a higher chance of showing early saturation and negative differential resistance. Quantum transport simulations show that the transfer length of Ni-MoS$_2$ contacts can be as short as 5 nm. Our results suggest that charge injection at the source contact differs from injection at the drain side: scaled source contacts can limit the drain current, whereas scaled drain contacts cannot. Furthermore, we clearly identified that the transfer length depends on the quality of the metal-2D interface. The asymmetrical contact measurements demonstrated here will enable further understanding of contact scaling behavior at various interfaces.

## 1. Introduction

Metal-semiconductor contacts are a fundamental building block for semiconductor devices and have been investigated for almost 150 years. In 1874, the first metal-semiconductor contacts were discovered, and their rectification effects were observed.[1] Then, in the 1930s, the theories on the rectification effects were developed to form the basis for understanding these contacts.[2–6] With the invention of the transistor[7], the development of ohmic contacts has been a major research topic in advancing various transistor technologies.[8,9] During the decades of research and especially with the development of different measurement methods (e.g., the transmission line method),[10] it was expected that charge carriers crowd near the tip of planar metal contacts, i.e., only a fraction of the contact is employed for active carrier injections. This current-crowding phenomenon significantly affects how small the contacts can be scaled without limiting drain currents. Recently, one of the significant frontiers for studying the metal-semiconductor contacts is their scaling behavior (i.e., how reducing device sizes impacts device operation and performance), manifested mainly in two trends. First, the thickness of semiconductor channel materials has been reduced with the advent of atomically thin semiconductors[11–14] and the development of gate-all-around nanosheet transistors.[15] This trend is called channel thickness scaling. Second, the metal contact length ($L_c$), the length where metal contacts overlap semiconductors (**Figure 1**a), needs to be scaled, termed contact scaling. These two trends are crucial as a thinner transistor channel produces better electrostatic control,[16] and smaller $L_c$ yields smaller device footprints and thus higher device densities in integrated circuits.[12] At the intersection of these two trends is the investigation of contact scaling on two-dimensional (2D) nanomaterials.

As illustrated in Figure 1b, the transfer length ($L_T$) represents the length across which most of the carriers are injected as the current path is the smallest combination of vertical specific contact resistivity ($\rho_c$) and lateral sheet resistance ($R_{sh}$) at the metal-2D material junction.[10] As $L_c$ approaches the transfer length, the contact resistance ($R_c$) is projected to increase due to current crowding[10]. Injected carriers in 2D FETs must tunnel through a physical van der Waals (vdW) gap in addition to a large Schottky barrier from strong Fermi level pinning.[17,18] Therefore, determining $L_T$ is key for assessing the scaling limits of 2D FETs and understanding the nature of metal-2D semiconductor interfaces.

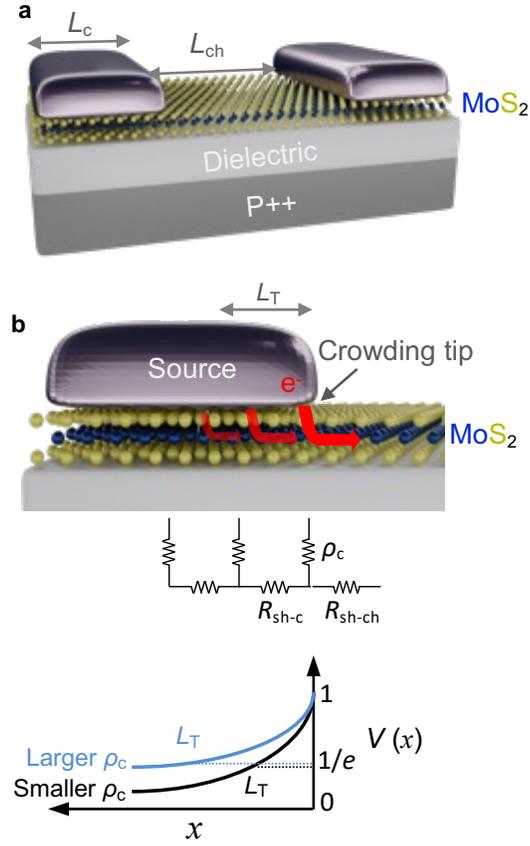

**Figure 1.** Representation of transfer length in top metal-2D contacts. (a) Diagram of a typical top-contacted, back-gated MoS₂ transistor with contact length ($L_c$) and channel length ($L_{ch}$) labeled. (b) Magnification of the source contact showing electrons injecting via the tip of the top contact. Annotations in the schematic oxide layer include the specific contact resistivity ($\rho_c$, unit: Ω·cm²) and sheet resistance (unit: Ω/square) underneath the contact ($R_{sh-c}$) and in the channel ($R_{sh-ch}$). Aligned below the diagram is the normalized potential $V(x)$ under the contact versus position ($x$) as a function of $\rho_c$. Transfer length, $L_T$, is the length where the potential drops to "1/$e$" of the original value (≈36.8%), where $e$ is the Euler's number.

The scaling behavior of metal-2D semiconductor contacts has been largely neglected. Most studies in contacting 2D materials report $L_c$ dimensions from hundreds of nanometers to a few micrometers.[12,13] Although a few studies investigated contact scaling for 2D FETs, incongruent results have been presented, as summarized in Table 1. For well-investigated transition metal dichalcogenides (TMDs) such as MoS₂, researchers reported a wide range of $L_T$.[19–24] Meanwhile, we also cannot find agreement on whether carrier density ($n$) in the channel impacts $L_T$. The inconsistency of $L_T$ values seen in Table 1 could stem from these studies' analytical approach. $L_T$ is often indirectly estimated from techniques such as the transfer length method (TLM), which ignores non-idealities at the metal-TMD interface (*e.g.*, non-uniform $\rho_{sh}$ underneath the contacts versus in the channel) and the contact-gating effect.[25]

Table 1. Summary of Contact Scaling Studies on MoS$_2$ FETs

| Contact metal | $T_{ch}$ | Depend on $n$? | $L_T$ (nm) | Ref. |
|---|---|---|---|---|
| Au | 6L | No | ≈40 | Exp.[19] |
| Au | 1L | No | ≈80 | Exp.[20] |
| Ti, Au | 2,6L | Yes | 20~200 | Exp.[21] |
| Ag | 2L | Yes | 114~128 | Exp.[22] |
| Ti/Au | 2L | Yes | 630 | Exp.[23] |
| Ni | 3L | No | <13 | Exp.[24] |
| Au | 1,2L | Yes | 1~2 | Sim.[26] |
| Ti | 1L | - | <1* | Sim.[27] |

Note: 1L: monolayer; 2L: bilayers, etc. *n*: carrier density in the channel
Exp.: experimental study; Sim.: theoretical simulation.
*The "$L_T$ < 1 nm" result was simulated by assuming a clean Ti-MoS$_2$ interface in Ref. [27].

The dichotomy between experimental and theoretical results is also apparent. While experimental results generally show a certain length for $L_T$, theoretical simulations, such as a semi-classical model[26] and *ab initio* quantum transport,[27] find carriers are injected via the tip of the top contacts for monolayer MoS$_2$ (essentially, $L_T$ is close to 1 nm). This discrepancy between experimental and theoretical studies further highlights the complexity of metal-2D interfaces. Hence, these inconsistencies necessitate a systematic investigation of the scaling behavior of metal-2D contacts based on new methods. Our study aims to answer the following fundamental questions: (1) how long is the transfer length in metal-2D interfaces? (2) how does the drain-source voltage ($V_{DS}$) impact transfer length? (3) is current crowding at the source and drain the same? Exploring these questions will help illuminate current crowding mechanisms at the metal-2D interface.

This work combines physically reduced contact length with asymmetrical contact measurements to explore contact scaling behavior in metal-2D contacts. Although $L_c \approx 13$ nm has been demonstrated,[24] the variation and the scaling behavior in the saturation regime remain unexplored. Asymmetrical contacts formed with different source and drain materials have been implemented, but mostly in applications such as photodetectors[28–30] and rectifiers.[31] Also, no asymmetrical contact lengths have been reported. Hence, it is intriguing to use asymmetrical contacts with different contact lengths to directly probe the contact scaling effects.

Critically, in our asymmetrical contact measurements, the same device is tested twice, first with the left contact as the source and then the right contact as the source. The resulting device characteristics highlight the comparison of different contact lengths as the source, while the channel remains unchanged. This approach eliminates the channel-to-channel variation, which often makes comparing emerging devices unviable and challenging. We first study gradual contact scaling, where we fabricate an array of devices on the same 2D film with gradually decreasing $L_c$, and find that scaled contacts (< 40 nm) can exhibit early saturation compared to longer contacts. To further investigate this emergent phenomenon, contacts with a larger difference of $L_c$ are utilized with adjacent symmetrical and asymmetrical contacts. We find that, while the two contacts perform similarly at small $V_{DS}$ in the linear regime, their characteristics are different in the saturation regime at large $V_{DS}$. This observation suggests that scaled contact length at the *source* electrode can yield degraded device performance. In comparison, a short contact length for the drain electrode has a negligible impact on the device performance, indicating different tolerances toward contact scaling for source and drain contacts. Lastly, we expand on asymmetrical contact measurements, illustrating that it is a valuable measurement technique for contact and interface engineering.

## 2. Results and Discussion
### 2.1 Gradual contact scaling

Back-gated MoS$_2$ FET test structures were fabricated with Ni contacts that have decreasing contact lengths from 120 nm to 30 nm deposited on the same bilayer MoS$_2$ (Figure 2). The bottom gate dielectric is 20 nm AlO$_x$ grown by using atomic layer deposition. The capacitance is measured to be around 280 nF/cm$^2$. In Devices 1 to 7 (D1-7), the channel length is designed to be the same while the contact length is varied. Both scanning electron microscopy (SEM) and scanning transmission electron microscopy (STEM) are used to confirm the contact lengths (≈ 120 nm, ≈ 105 nm, ≈ 85 nm, ≈ 65 nm, ≈ 45 nm, ≈ 40 nm, ≈ 35 nm, and ≈ 30 nm from left to right in **Figure 2**a). Example SEM and STEM images are shown in Figure 2b. The shape of the Ni contacts in STEM images does not depend on the contact length, which was thoroughly examined in a recent study.[32] Figure 2c plots the drain current as a function of the gate voltage ($I_D$-$V_{GS}$ characteristics) of D1-7. Because these devices have different threshold voltages ($V_T$) in the raw data, we align the $I_D$-$V_{GS}$ data such that $V_T$ is approximately 1.2 V. This way, the devices have the same overdrive voltage ($V_{OV} = V_{GS} - V_T$), and the curves are easier to compare. D1-7 with different $L_c$ have surprisingly similar performance at $V_{DS} = 1$ V. Because the seven devices have approximately the same channel resistance (same channel length), Figure 2c also indicates that the devices with different $L_c$ have similar contact resistance. The original $I_D$-$V_{GS}$ curves before $V_T$ alignment are given in Figure S1. Figure 2d confirms the total resistance, $R_{tot}$, (which consists of both the channel resistance and the contact resistances), is similar for devices with different contact lengths when they have comparable carrier densities. These results indicate that the transfer length can be smaller than 30 nm, the shortest $L_c$ measured here if the contact interface is clean.

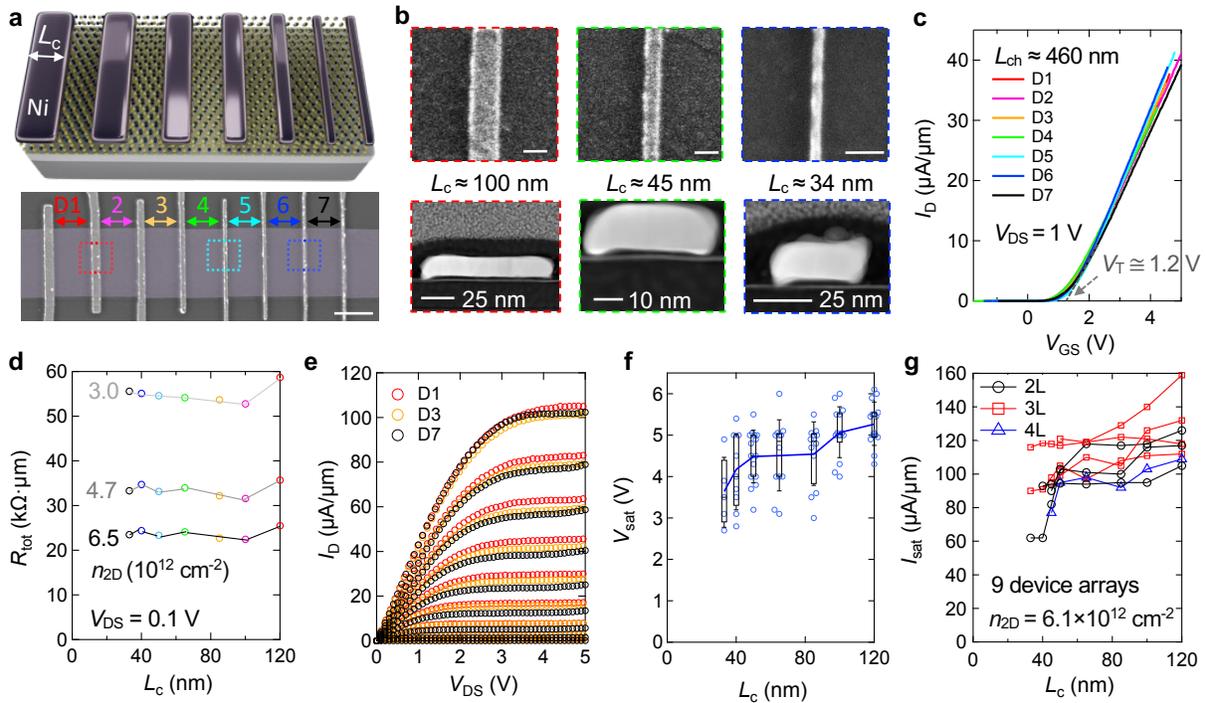

**Figure 2.** Gradual contact scaling with $L_c$ ranges from 120 nm to 30 nm. (a) Schematics and SEM image of the 2L MoS$_2$ devices with continuously decreasing $L_c$. From left to right, the $L_c$ is approximately 120 nm, 105 nm, 85 nm, 65 nm, 45 nm, 40 nm, 35 nm and 30 nm. Scale bar of the SEM image, 500 nm. (b) Higher magnification SEM and cross-sectional STEM images of the Ni contacts. Scale bars in the SEM images, 100 nm. (c) $I_D$-$V_{GS}$ for devices D1-7 at $V_{DS} = 1$V. The curves are aligned to have the same $V_T$ ($\approx 1.3$ V). (d) The total resistance ($R_{tot}$) is calculated for D1-7 under different carrier densities (3.0, 4.7, and $6.5 \times 10^{12}$ cm$^{-2}$). $R_{tot}$ is obtained at $V_{DS} = 0.1$ V. (e) $I_D - V_{DS}$ curves of three example devices. The D7 shows signs of early saturation. The $V_{DS}$ sweep direction is forward. $V_{GS}$ is from 7 V to 3 V in a stepwise of –0.5 V. (f) Boxplot of $V_{sat}$ from 9 arrays of devices with different $L_c$ as the source contact. (g) Saturation $I_D$ ($I_{sat}$) at $V_{DS} = V_{sat}$ and $n_{2D} = 6.1 \times 10^{12}$ cm$^{-2}$, as a function of contact length. 2L, 3L, and 4L represent the layer number of the MoS$_2$ channel materials.

To illustrate the full device characteristics, we show the $I_D$-$V_{DS}$ device characteristics in both the linear and saturation regimes (Figure 2e for Device 1, 3, and 7 as examples). These data show similar performance with different contact lengths when $V_{DS}$ < 1 V. But after $V_{DS}$ exceeds 1 V, the devices with smaller $L_c$ tend to saturate at lower $V_{DS}$ relative to those with longer $L_c$. To more clearly illustrate this trend, we show statistical data of the saturation voltage ($V_{sat}$) from 9 arrays of devices with different contact lengths in Figure 2f. The $V_{sat}$ is defined as the voltage at which the $I_D$ starts to saturate in the output characteristics. Figure 2f shows the $V_{sat}$ decreases for devices with contact lengths shorter than 80 nm. Also, the variation of the $V_{sat}$, represented by the standard deviation, noticeably increases from 0.5 to 0.9 as the contact length approaches ~30 nm (Figure S2). This observation highlights the impact of contact scaling on critical device characteristics, i.e., increased variation and increased probability of saturation at smaller $V_{DS}$ (early saturation).

Figure 1g shows the $I_{sat}$ vs $L_c$, at $V_{sat}$ and $n_{2D}$=6.1×10$^{12}$ cm$^{-2}$ (adjusted to $V_{GS}$=6.5 V, $V_T$=3 V) from the 9 device arrays. The data are separated by the MoS$_2$ thickness in the channel. From Figure 1g, 2 device arrays with 3L MoS$_2$ channel demonstrate little contact length dependence as $L_c$ decreases from 120 nm to 33 nm. However, the general trend is that shorter contact lengths degrade the $I_{sat}$ in all device arrays with the MoS$_2$ thickness ranging from 2L, 3L, to 4L. This trend is more pronounced when $L_c$ is shorter than 50 nm. These data indicate the current crowding scenario in the short contacts depends on the local contact quality and cleanliness.

## 2.2 Symmetrical-asymmetrical contacts

To further highlight the contrast between short and long contact lengths, we designed and fabricated devices with symmetrical and asymmetrical contacts on the same MoS$_2$ film (**Figure 3**a). Importantly, each device was characterized twice: once with the left contact as the source and again with the right contact as the source (Figure S3). For simplicity, we analyze the three devices in Figure 3a: A (symmetrical contacts with $L_c$ ≈ 100 nm), B (asymmetrical contacts with $L_c$ ≈ 100 nm on one side and $L_c$ ≈ 30 nm on another), and C (symmetrical contacts with $L_c$ ≈ 30 nm). As expected, no significant differences can be observed when switching sources in device A (Figure S4). Correspondingly, devices A and B perform very similarly when the source $L_c$ is ≈ 100 nm at both low and high $V_{DS}$ values (Fig. 3b). This comparison suggests that the short drain $L_c$ in device B is not a limiting factor for the device performance. However, in Figure 3c, different $V_{DS}$ values significantly impact $I_D$ in the asymmetrical contacts of device B. At $V_{DS}$ = 1 V (linear regime), the source with different $L_c$ yields the same *I-V* curves, suggesting similar contact and channel resistances. In contrast, at $V_{DS}$ = 4 V (in the saturation regime), with the shorter contact as the source, we observe a noticeable drop in $I_D$ in Device B, under the same overdrive voltage, $V_{OV}$.

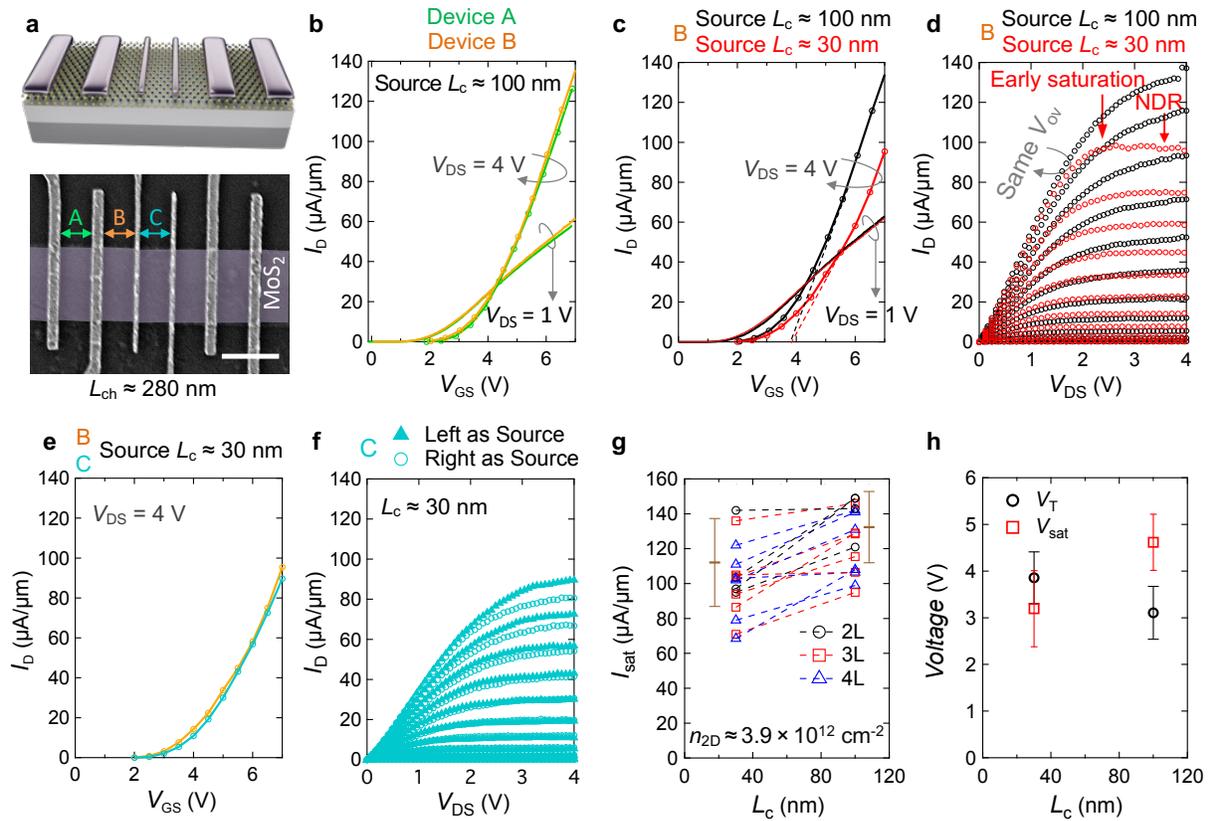

**Figure 3**. Devices with symmetrical-asymmetrical contacts show contact scaling effects at high $V_{DS}$. (a) Schematics and SEM of the devices. (b) Comparison of devices A and B at $V_{DS}$ = 1 V and 4 V. Both devices use $L_c \cong 100$ nm as the source. These data are the original curves without $V_T$ alignment. $I_D$-$V_{GS}$ (c) and $I_D$-$V_{DS}$ (d) characteristics of device B with different $L_c$ as source. In (d), the $V_{ov}$ is from 3.2 V to −1.2 V with a stepwise of −0.5 V. (e) Comparison of $I_D$-$V_{GS}$ between asymmetrical (B) and symmetrical (C) contacts with the same $L_c$ as source at $V_{DS}$ = 4 V. (f) $I_D$-$V_{DS}$ of device C with left or right contacts as source. In (e-f), the $V_{ov}$ is from 3 V to -1 V with a stepwise of −0.5 V. (g) Statistics of $I_{sat}$ vs $L_c$ from 17 asymmetrical devices. The mean values are represented in the offset brown data. (h) $V_T$ and $V_{sat}$ of 17 devices with different $L_c$ as the source contact. The error bars in (g-h) represent one standard deviation from the mean value.

A complete representation of the impact of $V_{DS}$ is demonstrated in the $I_D$-$V_{DS}$ curves in Figure 3d. In the linear regime, sources with different contact lengths produce similar $I_D$. In contrast, in the saturation regime ($V_{DS}$ = 4 V), smaller contact lengths as the source have decreased $I_D$, confirming the observation in Figure 3c. Surprisingly, we also observe apparent early saturation and negative differential resistance (NDR) when $V_{DS}$ increases from 2 to 5 V for the short contact length as the source. This NDR and early saturation are repeatable for other asymmetrical devices (Figure S5). Although the NDR is not observed in device C with symmetric short contacts (Figures 3e,f), around 33% decrease in $I_{sat}$ is observed relative to the source $L_c$ = 100 nm devices.

The statistical distributions of $I_D$ at $V_{DS}$ = 4V vs. $L_c$ from the 17 asymmetrical devices are shown in Figure 3g. On the same exact device, the average $I_{sat}$ values decrease by ~15% for 30 nm contacts compared to 100 nm contacts, indicative of the contact scaling effect driven by scaled contact length at the source electrode. However, 4 of the 17 devices demonstrate relatively consistent $I_{sat}$ values regardless of the source contact lengths, showing the possibility that the transfer length can be much smaller than 30 nm. These 4 devices include devices based on 2L, 3L, and 4L MoS$_2$, which suggest that if the contact interface is clean, the transfer length can be small for channel thicknesses ranging from 2L to 4L. $V_{sat}$ and $V_T$ vs. $L_c$ from the 17 asymmetrical devices are shown in Figure 3g. With $L_c \cong$ 30 nm contact as the source, $V_{sat}$ decreases by ~30% (from 4.6 V to 3.2 V) on average, representing early saturation induced by scaled source contacts. With 100 nm contact length as the source, the $V_T$ on average is 3.9 V, and the $V_T$ decreased to 3.1 V when using the 30 nm contact as the source. Usually, $V_T$ shift is related to the charge trapping in the channel. We expect that switching source and drain can cause the charge trapping configuration to be different, thus introducing $V_T$ differences. However, from Figure S6, the $V_T$ shift from symmetrical devices with $L_c \cong$ 100 nm is quite small. Hence, the $V_T$ shift in asymmetrical devices can be partially attributed to the short source contact length. The detailed mechanism merits future independent investigation.

Our observations in Figure 3 are not related to the breakdown of the channel near the drain side. The breakdown of the monolayer MoS$_2$ has been observed near the drain contacts under high electric fields from drain to source.[36] One example device fabricated with the same device geometry as Device A in Figure 3 partially breakdown at $E_{DS}$ = 21 V/μm (**Figure 4**a). Although this breakdown appears similar to the breakdowns near the drain contact observed in Smithe *et al.*,[36] the devices in Figure 3d saturate at $E_{DS}$ = 10.3 V/μm, which is only half of the $E_{DS}$ needed to induce partial breakdown. This comparison suggests that the device in Figure 3d does not reach the breakdown state. Also, no depletion region forms near the drain since $V_{DS} < V_{GS} - V_T$ in Figure 3d. Moreover, the $I_{sat}$ in Figure 4a is two times higher than the $I_{sat}$ value in Figure 3d, indicating high $I_{sat}$ is needed to induce partial breakdown. These comparisons confirm that the NDR and early saturation can be attributed to the short $L_c$ in the source, not to hot electron effect-induced breakdown near the drain side.

Our observations in Figure 3d are not related to the $I_D$-$V_{DS}$ sweeping directions. Forward sweeping can induce early saturation and NDR compared to backward sweeping, as shown in Figure 4b. However, the comparisons in Figure 3d are all on the forward sweep, which should yield the same behavior if current crowding happens at the tip of the top contacts for the source. Also, the early saturation in Figure 4b happens at $V_{DS} = 4$ V, whereas the early saturation in Figure 3d occurs at $V_{DS} = 2$ V (on average, at $V_{DS} = 3.1$ V from Figure 3g). The different drain-to-source fields at which early saturation occurs underline the impact of $L_c$ of the source, i.e., the smaller the source $L_c$, the early saturation tends to happen at smaller $V_{DS}$. Finally, although the forward sweep in Figure 4b has early saturation and NDR, the $I_{sat}$ is still on par with Device A in Figure 3a. This comparison corroborates the earlier observation that short source $L_c$ can cause reduced $I_{sat}$.

Hence, the asymmetrical scaling behavior in Figure 3 is caused by the small source contact length. Specifically, as the contact length decreases, more electrons can be injected from the metal to the $MoS_2$ but are limited by the available contact length. However, when the contact length is large, the carriers are injected via all the available contact interface areas, thus having larger $I_{sat}$ and $V_{sat}$. We have observed a few device arrays with consistent $I_{sat}$ regardless of the $L_c$ (larger than 30 nm). Hence, the transfer length in a specific device can be smaller than 30 nm. From Figure 3g, smaller contact length also induces more variation. From these observations, we have gained a deeper understanding of current crowding at the metal-2D interface: the transfer length should depend on the cleanliness and quality of the metal-2D interface. When the Ni-$MoS_2$ interface is less ideal or unclean, the transfer length is larger, and more contact length limited transport can be observed.

NDR can be induced by many mechanisms[33] and has been observed in multiple 2D FETs.[34,35] Several possible mechanisms are evaluated for the decreased saturation voltage and increased possibility of NDR at shorter source $L_c$. First, these observations are unlikely to be caused by the arched shape of the Ni contacts. From Figure 2b and our recent study of metal/2D contact geometry,[32] the arched shape of Ni contacts is present for both short and long contact lengths, thus not likely to be responsible for behaviors observed in Figure 3.

On the other hand, NDR in $MoS_2$ FETs is commonly attributed to self-heating.[36–38] The NDR in Figure 3d is not related to self-heating for two reasons: first, self-heating typically causes decreased transconductance at the saturation regime, which is not observed in Figure 3d; second, self-heating typically happens for high drain current devices. Yet, in Figure 3d, it is the smaller $I_D$ that exhibits the NDR. We also benchmarked the reported NDR among MoS$_2$ FETs in Figure 4c. Compared to room-temperature, low-temperature measurements tend to induce NDR due to increased $I_{sat}$.[37] Compared to self-heating induced NDR in other devices, the contact-scaling induced NDR occurs at a smaller drain-to-source field ($E_{DS}$) and carrier density (representing the vertical field). Although different devices might have different channel thicknesses, contact metals, and oxide materials, this comparison still highlights the impact of contact scaling on the saturation regime operation.

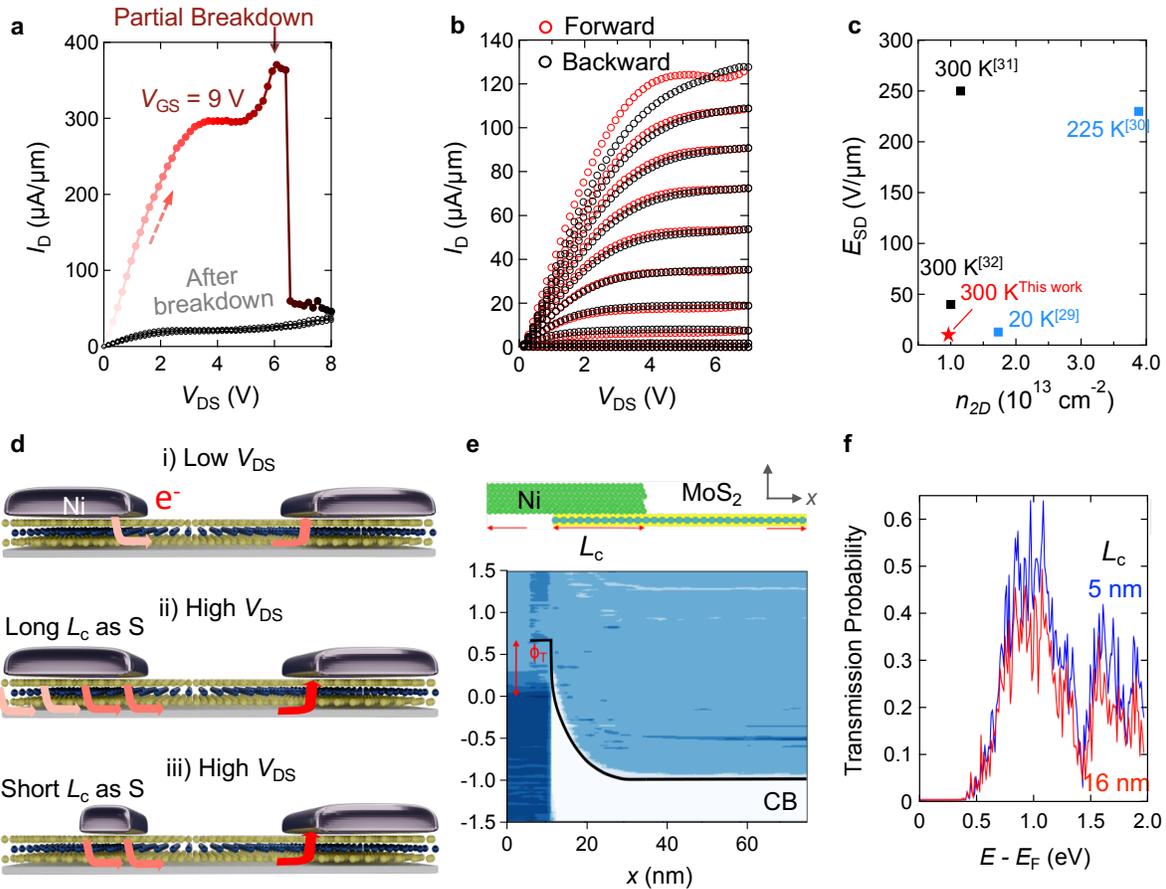

**Figure 4**. Mechanism of electron injection at low and high $V_{DS}$. (a) Partial breakdown of a MoS$_2$ transistor with $L_{ch}$ of ≈280 nm. (b) Forward and backward $I_D$-$V_{DS}$ sweeps of a Ni-contacted MoS$_2$ transistor with the same device dimensions as Device A of Figure 3a. $V_{OV}$ ranges from 7 V to -1 V in a step size of 1 V. (c) Benchmarking of NDR in MoS$_2$ FETs under different measurement temperatures. Low-temperature measurements are denoted in light blue. (d) Electron injection at a scaled contact interface at low and high $V_{DS}$. At high $V_{DS}$, a longer transfer length is expected and electron injection happens at a larger portion of the contact length. The short contact length becomes a limiting factor for carrier injection. (e) Quantum simulation of Ni-MoS$_2$ top contacts with different contact lengths. The schematic of the simulated interface is shown with the middle red arrow representing the contact length. Below the schematic, the spatially resolved density-of-states (DOS) is shown where darker

color indicates higher DOS. The black line denotes the MoS$_2$ conduction band (CB) minima. (f) Transmission probabilities at the Ni-MoS$_2$ top contact interface with 5 nm and 16 nm contact lengths.

Based on the data in Figure 3, Figure 4d depicts the mechanism contributing to the early saturation and NDR. At low $V_{DS}$ (linear regime), the carriers are injected through the tip of the top metal-2D interface, which explains the nearly identical *I-V* characteristics at low $V_{DS}$ for both small $L_c$ and longer $L_c$. However, at high $V_{DS}$ (saturation regime), when the metal-2D interface is not clean or specific contact resistivity is large, a larger portion of the contact will participate in the carrier injection. If the source contact is scaled, it becomes the limiting point for the number of electrons injected into the channel. The lower number of electrons injected decreases $I_{sat}$ and produces early saturation. As $V_{DS}$ further increases, the short source $L_c$ further limits the carrier injection, increasing the probability of NDR.

After the electrons are injected to the channel and accelerated by the high $V_{DS}$, these high-energy electrons are injected more easily into the drain side metal via the tip of the metal-2D contacts. In other words, the drain contact length does not make a difference in the range of $L_c >$ 30 nm, which also explains why the *I-V* characteristics are nearly the same in Figure 3a. Essentially, Figure 4d suggests that the source and drain can have different contact scaling behavior. Hence, the transfer length for source and drain can also be different.

To investigate the impact of top contact length scaling, we simulated the quantum transport dynamics at the metal-2D interface based on a non-equilibrium Green's function (NEGF) approach (Note S1). Figure 4e shows the schematics of the simulated Ni-MoS$_2$ interface. The simulation sets the overlap length between the metal atoms and the 2D crystal as 5 nm and 16 nm. In the spatially resolved density-of-states (DOS) plot, the conduction band minimum of the MoS$_2$ was pushed down 1 eV below the metal Fermi level, representing the electrostatic potential applied from the bottom gate. The resulting Schottky barrier height ($\phi_T$) is calculated to be 0.66 eV. Figure 4f also depicts the simulated transmission probabilities of the 5 nm and 16 nm contact lengths, showing a similar transmission probability profile. At higher energies, the 5 nm contact length does have a slightly higher transmission probability. We attribute this counter-intuitive observation to the interference between electrons in the overlap region for the longer overlap length (16 nm), similar to waves in a cavity. This result suggests the transfer length for the Ni-MoS$_2$ interface in principle can be as short as 5 nm, in agreement with earlier simulations of clean Ti-MoS$_2$ contact interfaces.[27] The simulation does not include dissipative and thermal effects such as self-heating.

Also, the cleanliness of the interface is not considered and it likely affects the transfer length. A detailed description of the quantum simulation and the comparison of Ni-MoS$_2$ top and edge contacts are given in Note S1.

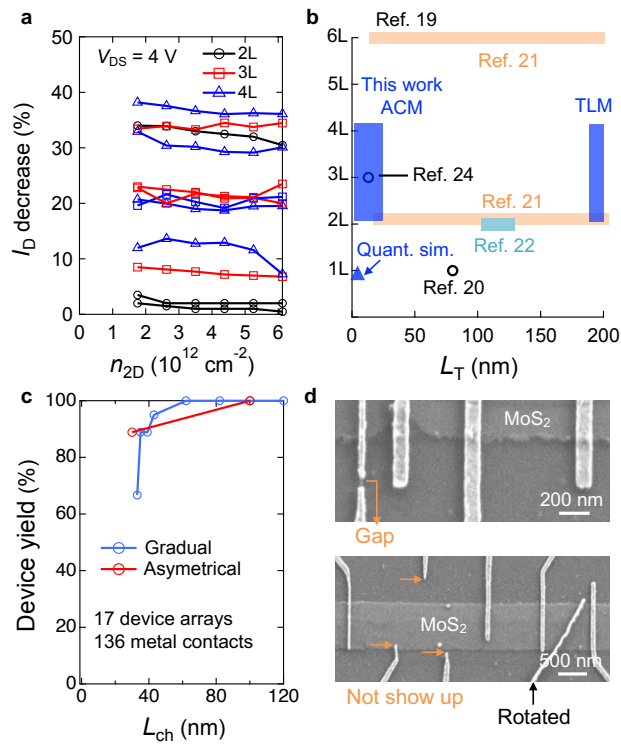

**Figure 5**. The impact of carrier density, transfer length benchmarking and device yield of scaled contacts. (a) The percentage of $I_{sat}$ drop versus the carrier densities of in the channel. (b) Benchmarking the transfer length range between the literature. (c) Device yield comparison between scaled and long contacts. (d) Representative failed modes of scaled contacts.

How do carrier densities impact the $I_{sat}$ in the asymmetrical contact comparison? After aligning the $V_T$ of the asymmetrical devices, the carrier density $n_{2D}$ can be approximated as $C_{ox}(V_{GS}-V_T)/q$, where $q$ is the elementary charge and $C_{ox}$ is about 280 nF/cm$^2$. As shown in Figure 5a, although the short contact as source degrades at a different percentage in different device sets, no trend can be established from the impact of carrier densities. This observation is in contrast with the conclusion demonstrated in Ref. We note that in their study, the dependence of carrier densities comes from analytical methods. Hence, our measurements provide direct experimental evidence for the impact of carrier density on contact scaling behavior.

To benchmark $L_T$ with the literature, our measurements of gradual contact scaling and asymmetrical contact scaling suggest an approximate range of $L_T <$ 30 nm, as depicted in Figure 5b. More significantly, we have shown that the $L_T$ range likely depends on the local metal-2D interface, which can be a source of variation. We note that the $L_T$ < 30 nm range does not denounce the possibility of $L_T \approx 13$ nm demonstrated in Ref. [24]. However, we note two key distinctions. First, the conclusion of $L_T \approx 13$ nm in Ref. [24] is based on devices with different channels, whereas the asymmetrical devices in our study permit the different contact lengths sharing the same exact channel, thus decreasing variation for comparison. Second, when plotting $I_D$ vs $L_c$ at $V_{DS}$ = 1 V for $L_{ch}$ = 29 nm and $L_{ch}$ = 500 nm, the authors find similar $I_D$ spread for device with $L_{ch}$ = 29 nm (saturation regime) and $L_{ch}$ = 500 nm (linear regime). In contrast, in our gradual scaling and asymmetrical contact measurements, devices perform similarly in the linear regime and start to differ in the saturation regime. Our results and the measurements thus promise to unveil unique contact scaling phenomena in different operation regimes.

It is constructive to gain insight on the device yield of scaled contacts. We use *I-V* measurements to determine the device yield. If no drain current can be measured, the device is considered open and the metal contacts likely failed. Although there is a possibility that the metal contacts work and it is the channel material that is open, the chance of the continuous $MoS_2$ film suddenly disconnecting entirely at the channel is expected to be relatively low. In Figure 5c, metal contacts with $L_c$ of over 50 nm tend to yield high ($\cong$ 100%). But as $L_c$ drops below 40 nm, the yield starts to drop, reaching a range of 65 to 85% at $L_c \approx$ 30 nm. This yield range is attributed to experiment-to-experiment variation. We also highlight the different failure modes of scaled contacts, such as forming a gap, to not showing up, and rotation. Further research is warranted to improve the yield of scaled contacts.

Moving forward, knowing the exact limit of transfer length by fabricating even smaller scaled contacts ($L_c$ < 20 nm) is valuable. However, our study also underscores the critical need to focus on improved and cleaner metal-2D contact interfaces, which dominate the contact scaling behaviors and realistically determine the device variation. It is also worthwhile to investigate whether P-type 2D transistors have a similar response to scaled source contacts. Further experimental work to examine the frequency response will show whether pulse measurements can help eliminate the NDR and sustain the drain currents. Temperature dependence measurement on the asymmetrical contacts can also compare the Schottky barrier height for the contacts with different contact lengths. Hopefully, a deeper understanding of the transport mechanisms will allow the improvement of 2D devices with scaled contacts.

## 2.3 Asymmetrical Contacts Measurements (ACMs)

As demonstrated in Figures 2 and 3, test structures with asymmetrical contacts can be versatile and powerful tools to explore the contact scaling behavior of $MoS_2$ transistors. Here we expand this method and propose an extended set of structures that can be used for other 2D materials as well as other families of low-dimensional nanomaterials (1D, organic and oxide thin films, etc.). **Figure 6**a shows the structures employed in Figures 2 and 3 with various emerging channel materials. One possible concern is that channel-to-channel variations might impact the accuracy and reliability of measurements based on the structures in Figure 6a. To address this potential issue, we propose a device geometry with both symmetrical contacts and asymmetrical contacts sharing the same channel area in Figure 6b for non-1D materials. Assuming the channel material is isotropic, sharing the same channel region can further eliminate the channel-to-channel variability that is common for low-dimensional nanomaterials.

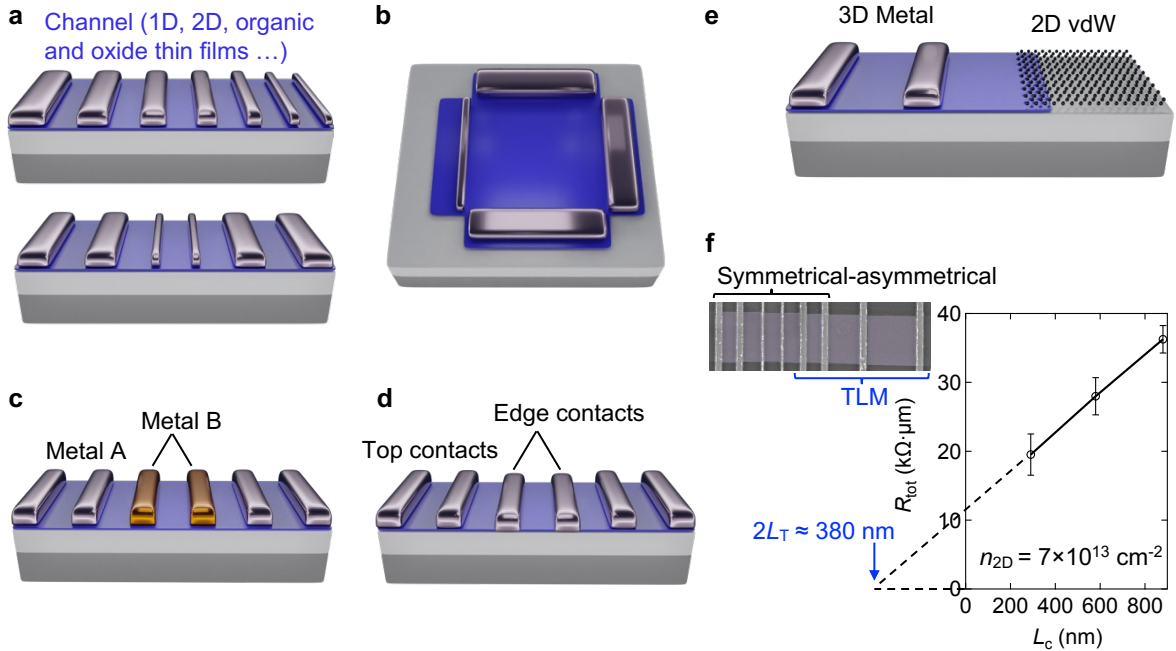

**Figure 6**. Example structures of asymmetrical contacts measurements (ACMs). (a) Gradual contact scaling and symmetrical-asymmetrical contact scaling for low-dimensional nanomaterials. (b) Symmetrical and asymmetrical contacts share the same channel. Symmetrical and asymmetrical contacts with different contact metals (c) and top vs. edge contacts (d). (e) Symmetrical devices with 3D metal contact and asymmetrical contacts with 3D metal and 2D vdW contacts. (f) Combining

asymmetrical contact measurements and TLM. The error bars represent one standard deviation from the mean value.

Additionally, different contact metals can exhibit different scaling behavior; hence, Figure 6c demonstrates a structure that enables measurements of three symmetrical contact devices as well as two asymmetrical devices with different metals as contacts. By scaling all contacts to the same length (for instance, below 50 nm), the different scaling behavior can unveil the impact of different metal properties (metal vs. semimetal, work functions, deposition methods/conditions, etc.). Apart from using different metals, edge contacts are a promising approach that may offer the ultimate immunity to contact scaling.[39] Therefore, the scaling behavior comparison between the top and edge contacts is particularly interesting to investigate (Figure 6d).[40] Finally, metallic or semi-metallic 2D vdW materials such as Graphene have been investigated as the contacts for $MoS_2$ FETs.[41,42] It is possible that such 2D vdW contacts have different scaling characteristics. Figure 6e illustrates a symmetrical-asymmetrical device structure that compares 3D metal and 2D vdW contacts.

Combining ACMs with traditional measurement approaches (TLM, Four-probe, etc.) is recommended to reveal further insights into the impact of scaling $L_c$ and other properties of the metal-semiconductor contact interface. Figure 6f depicts an example of fabricating ACMs and TLM test structures onto the same $MoS_2$ film. $L_T \approx 190$ nm is determined from the TLM measurement. The TLM-derived $L_T$ is much larger than the short $L_T$ range indicated by the contact scaling behavior observed in Figures 2 and 3. This comparison highlights the advantages of using ACMs to investigate contact scaling and current crowding effects. ACMs will expand the available toolbox to investigate various contact interfaces and their scaling behavior. The resulting understandings and insights will benefit myriad device applications besides transistors, such as photodetectors, memristors, and sensors.

## 3. Conclusions

The nature of the metal-semiconductor interface has fascinated scientists for nearly 150 years. The scaling behavior of metal-2D interfaces has major implications for extending the grand endeavor of building smaller transistors. We investigated the realistic and statistical contact scaling behavior of $MoS_2$ FETs by proposing and using ACMs. The experimental observation confirms that contacts with small contact lengths can have similar contact resistances to large contact lengths. However, our results highlight that shorter source contact lengths (< 40 nm) have a higher chance of NDR, early saturation, and reduced performance at high $V_{DS}$. The traditional transfer length measurement method for determining contact resistance and channel resistivity may not fully depict how shorter contacts can impact device operation. Looking forward, additional studies are merited to reduce the variability of scaled contacts and to further understand the impact of substrates and contact metals. ACMs are also useful platforms for studying contact scaling in other low-dimensional nanomaterials. Characterizing devices with ACMs at lower temperatures (< 300 K) could open a new window for exploring metal-semiconductor contacts. Finally, because some contacts in mature transistor technology nodes are not gated, it is essential to explore the effects of contact gating on the scalability of contacts in future studies.

## 4. Experimental Section

*CVD Growth of the MoS₂.* As previously reported[43], 1g of sulfur powder (Sigma-Aldrich) and 15-30 mg of MoO₃ (99.99%, Sigma-Aldrich) were placed upstream and at the center of a tube furnace. The substrates (heavily doped Si substrate with 300 nm SiO₂) were placed downstream in the furnace tube. The growth was performed at 750 °C for 10 minutes under a flow of argon gas at a rate of 100 sccm at ambient pressure.

*Device Fabrication.* The MoS₂ CVD film was transferred onto AlO$_x$/P++ Si substrate. Electron beam lithography (EBL) with Poly(methyl methacrylate) (PMMA) was first used to define an MoS₂ bar as the channel region followed by CF₄ plasma etching (PMMA also serves as the mask during the etching). Electron beam lithography (EBL) with PMMA was then used to define the contacts, leads, and pads with electron beam evaporation used to deposit the metals. 15 nm of Ni was evaporated for the contacts, whereas 5 nm Cr/50 nm Au was used for the leads and pads. The fabricated devices were then characterized in the probe station under low pressure (~$10^{-3}$ Torr).

*HAADF-STEM preparation and characterization.* An FEI Nova NanoLab 600 DualBeam (SEM/FIB) was employed to prepare cross-sectional STEM samples. An electron beam-induced Pt deposition around 100 nm was deposited over the device to protect the sample surface, followed by a 2 μm Pt deposition with ion beam. The STEM samples were thinned with 30 kV Ga-ions beam while final thinning was performed at 2 kV to reduce damage. The Z-contrast HAADF-STEM images were collected using an FEI Titan 80-300 probe-corrected STEM/TEM microscope operating at 300 keV, with a beam convergence angle of 20 mrad and collection angles > 50 mrad.

**Author Contributions**

Z.C. designed the experiments, fabricated and characterized the devices. Z.C. analyzed the data with the assistance of C.A.R., A.D.F. and H.A. H.Z. prepared the TEM samples and conducted the STEM characterization. G.Z, Y.Y. grew the $MoS_2$ using CVD. J.B. and M.L. performed the quantum transport simulation of the top contacts and compared the results with edge contacts. Z.C., C.A.R, and A.D.F. wrote the main paper and the supporting information with input from all other co-authors.

**Corresponding Authors**

Zhihui Cheng – orcid.org/ 0000-0003-1285-0523; Email: zhihui.cheng@alumni.duke.edu

Curt A. Richter – orcid.org/0000-0003-4510-1465; Email: curt.richter@nist.gov

Aaron D. Franklin – orcid.org/ 0000-0002-1128-9327; Email: aaron.franklin@duke.edu

**Conflict of Interest**

The authors declare no competing interests.

**Data Availability Statement**

Data are available from the first and corresponding author, Z.C., upon reasonable request.

**Notes**

Certain commercial equipment, instruments, or materials are identified in this paper in order to specify the experimental procedure adequately. Such identifications are not intended to imply recommendation or endorsement by the National Institute of Standards and Technology (NIST), nor it is intended to imply that the materials or equipment identified are necessarily the best available for the purpose.

**Funding**

Z.C. acknowledges financial support from Semiconductor Research Corporation (SRC) nCORE program sponsored by NIST through award number 70NANB17H041. Z.C., H.A. and A.D.F acknowledge support from the National Science Foundation (ECCS 1915814). H.Z. acknowledges support from the U.S. Department of Commerce, NIST under financial assistance award 70NANB19H138. A.V.D. acknowledges support from the Material Genome Initiative funding allocated to NIST. J.B and M.L acknowledge support from NCCR MARVEL, funded by the Swiss National Science Foundation (SNSF) under grant No. 182892, by the SNSF under grant No. 175479 (ABIME), and by the Swiss National Supercomputing Center (CSCS) under projects s876 and s1119.

**Acknowledgements**

Fabrication and measurements were partially performed at the NIST Center for Nanoscale Science and Technology and at Duke Shared Manufacturing and Instrument Facility (SMIF). The authors would like to thank the staff members in the Shared Instrument and Manufacturing Facilities (SMIF) at Duke for their assistance.

Note about reference [1]

For an English version, see S. M. Sze, Ed., Semiconductor Devices: Pioneering Papers. Singapore and Teaneck, NJ: World Scientific, 1991: "On current conduction through metallic sulfides", pp. 377-380.

Schottky, W. "Halbleitertheorie der Sperrsschicht." Naturwissenschaften Vol. 26 (1938) pp. 843.
Abstract in English as "Semiconductor Theory of the Blocking Layer" in Sze, S.M. Semiconductor Devices: Pioneering Papers. (World Scientific Publishing Co., 1991) pp. 381.